\documentclass{PoS}

\usepackage{bbold}
\usepackage{amsmath}
\usepackage{slashed}
\newcommand{\be}{\begin{equation}}
\newcommand{\ee}{\end{equation}}
\newcommand{\bea}{\begin{eqnarray}} 
\newcommand{\eea}{\end{eqnarray}}

\title{Perturbative investigation of Wilson-line operators in Parton Physics}

\ShortTitle{Perturbative investigation of Wilson-line operators}

\author{\speaker{Gregoris Spanoudes} and Haralambos Panagopoulos\\
        Department \ of \ Physics, \ University \ of \ Cyprus, \ POB \ 20537, \ 1678, \ Nicosia, \ Cyprus\\
        E-mail: \email{spanoudes.gregoris@ucy.ac.cy},
        \email{haris@ucy.ac.cy}}

\abstract{
\qquad We investigate the renormalization of a class of gauge-invariant nonlocal quark bilinear operators, including a finite-length Wilson-line (called Wilson-line operators). The matrix elements of these operators are involved in the recent ``quasi-distribution'' approach for computing light-cone distributions of Hadronic Physics on the lattice. We consider two classes of Wilson-line operators: straight-line and staple-shaped operators, which are related to the parton distribution functions (PDFs) and transverse momentum-dependent distributions (TMDs), respectively. We present our one-loop results for the conversion factors of straight-line operators between the RI$'$ (appropriate for nonperturbative renormalization on the lattice) and $\overline{MS}$ (typically used in phenomenology) renormalization schemes in the presence of nonzero quark masses. 

\qquad In addition, we present the first results of our preliminary work for the renormalization of staple-shaped operators both in continuum (Dimensional Regularization) and lattice (Wilson/clover fermions and Symanzik improved gluons) regularizations. We identify the observed mixing pairs among these operators, which must be disentangled in the nonperturbative investigations of heavy-quark quasi-PDFs and of light-quark quasi-TMDs. 
}

\FullConference{XIII Quark Confinement and the Hadron Spectrum - Confinement2018\\
		31 July - 6 August 2018\\
		Maynooth University, Ireland}

\begin{document}

\section{Introduction - Outline}

A direction of research which has seen rapid progress in recent years regards the nonperturbative study of light-cone partonic distribution functions on the lattice. These functions provide important information on the quark and gluon structure of hadrons; at leading twist, they give the probability of finding a specific parton (quark, antiquark, or gluon) in the hadron carrying certain momentum and spin, in the infinite momentum frame. The direct computation of parton distributions on the lattice has been made feasible only recently  by X. Ji's approach \cite{Ji:2013dva}. The application of this method is currently under investigation by many research groups and so far the outcomes are very promising for the correct estimate of a physical light-cone distribution function. In this approach, nonlocal fermion bilinear operators, which contain a finite-length Wilson line (called Wilson-line operators), are involved. The investigation of the perturbative renormalization of these operators is the goal of our study.

The outline of this paper is as follows. In Sec. 2 we provide a brief introduction to the Wilson-line operators and their relation to the parton distribution functions, as well as a short description of the quasi-distribution approach in Hadronic Physics and its application on the lattice. Sec. 3 contains the calculation setup and the results of our one-loop studies on the renormalization of two types of Wilson-line operators: the straight-line and the staple-shaped operators, which are relevant to the studies of parton distribution functions (PDFs) and transverse momentum-dependent distributions (TMDs) respectively. Finally in Sec. 4, we conclude with possible follow-up work.

\section{Wilson-line operators and PDFs}

\begin{figure}
\vspace{-0.1cm}
\centering
\includegraphics[height=1.5cm]{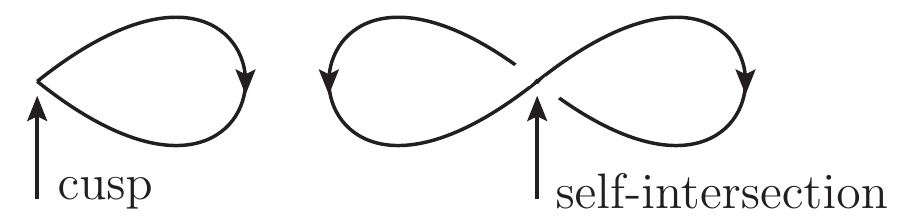}
\caption{Contours with singular points: cusps and self-intersections} 
\label{fig:Cusps}\vspace{-0.1cm}
\end{figure}

A Wilson line $W_C (x_i, x_f)$ is a phase factor given by the path-ordered exponential of a gauge field $A^\mu (x)$, along a line $C$ with end-points $x_i$ and $x_f$: 
\be 
W_C (x_i, x_f) = \mathcal{P} \Big\{ \exp \Big[i g \int_C dx_\mu A^\mu (x)\Big]\Big\}
\ee
This object has two important features: it is nonlocal and gauge-covariant. Quantities with these traits are very promising for the investigation of nonperturbative phenomena, such as quark confinement. The history of studies of operators involving Wilson lines in gauge theories goes back a long time, including  seminal work of Mandelstam, Polyakov, Nambu, Gervais - Neveu, Makeenko - Migdal, Witten, and many others. Some prominent work considers the renormalization of closed Wilson-loop operators \cite{Dotsenko:1979wb}: $\langle tr ( \mathcal{P} \{ \exp [i g \oint_C dx_\mu A^\mu (x)]\})\rangle$, which are related to the elementary excitations of a gauge field. In these studies, it was shown that, in regularizations other than dimensional regularization (DR), linear divergences arise even in smooth (i.e., differentiable) and simple (i.e., non-self-intersecting) contours. These divergences can be eliminated to all orders in perturbation theory, by an exponential renormalization factor that depends on the length $L$ of the contour and the ultraviolet cut-off scale $a$: $Z \sim \exp (- c \ L/a)$, where $c$ is a dimensionless quantity. Also, contours containing singular points \cite{Brandt:1981kf}, such as cusps and self-intersections (Fig. \ref{fig:Cusps}), introduce additional multiplicative renormalization factors. 

Another work, relevant to our study, is the renormalization of Wilson-line bilocal fermion operators \cite{Craigie:1980qs}: $\bar{\psi} (x) W_C (x,y) \psi (y)$, which are associated with the construction of gauge-invariant hadronic bound states. In these calculations, additional divergences, compared to those of the Wilson loops, arise from the end points of the open Wilson line. The insertion of a Dirac matrix in the definition of the Wilson-line bilocal fermion operators, leads to the construction of a composite operator $\mathcal{O}_\Gamma (x)$, involved in the definition of quark PDFs:
\be  
q_{_\Gamma} (x) = \int_{- \infty}^{\infty} \frac{d \xi^{-} }{4 \pi} \exp (-i x P^+ \xi^-) \Big\langle N \Big\vert \mathcal{O}_\Gamma^{\xi^-} (x) \Big\vert N \Big\rangle, 
\ee
where
\be
\mathcal{O}_\Gamma^{\xi^-} (x) = \overline{\psi} (\xi^-) \ \Gamma \ W (\xi^{-}, 0) \psi (0), \quad
W (\xi^-, 0) = \mathcal{P} \Big\lbrace \exp \Big[-i g \int_0^{\xi^-} d\eta^- A^+ (\eta^-)\Big] \Big\rbrace,
\ee
$\Big \vert N \Big\rangle$ is a hadron state, $x P^+$ is the fraction of the hadron momentum $P^+$, carried by each consituent parton inside the hadron, and $\Gamma = \gamma^+,\gamma_5 \gamma^+,\gamma^+ \gamma^\perp$ corresponds to the unpolarized, helicity and transversity PDFs respectively. PDFs are an essential tool for the analysis of Deep Inelastic Scattering (DIS) experiments, as they give the contribution of soft gluon radiation in the scattering. Because of the highly nonlinear nature of the parton dynamics for small values of x, PDFs can be evaluated only with nonperturbative methods. Lattice QCD is the only known nonperturbative theoretical framework, in which one can calculate PDFs from first principles. However, the involved Wilson line in the definition of PDFs, is defined in the light cone $({(\eta^-)}^2 = 0)$ and thus PDFs cannot be computed directly on a Euclidean lattice.   

A few years ago, a pioneering method for the direct computation of PDFs on the lattice was suggested by X. Ji \cite{Ji:2013dva}. This approach is summarized in three steps:
\begin{enumerate}
\item First, instead of computing light-cone correlation functions, one projects outside of the light cone and considers equal-time correlation functions, which are called quasi-distribution functions. For example, the definition of parton quasi-distribution functions (quasi-PDFs) is: 
\be 
\widetilde{q}_{_\Gamma} (x, \mu, P_\mu) = \int_{- \infty}^{\infty} \frac{dz}{4 \pi} e^{-i x P_\mu z} \Big\langle N \Big\vert \overline{\psi} (z) \ \Gamma \ W(z,0) \psi (0) \Big\vert N \Big\rangle, 
\ee
where 
\be 
W(z,0) = \mathcal{P} \Big\lbrace \exp \Big[-i g \int_0^z d \zeta A_\mu (\zeta)\Big] \Big\rbrace
\ee
and $\Gamma = \gamma_\mu, \gamma_5 \gamma_\mu, \gamma_\mu \gamma_\perp$. These functions are purely Euclidean and thus they are accessible on the lattice. The Wilson line involved in quasi-PDFs is a straight line in a spatial direction $\mu$. 
\item The second step is the renormalization of quasi-distribution functions. Since these functions are calculable on the lattice, one can renormalize them nonperturbatively in the lattice regularization, using a Regularization-Independed (RI$'$)-like scheme \cite{Alexandrou:2017huk, Chen:2017mzz}. The lattice version of the straight Wilson line is given by $\displaystyle W(z,0) = {\Big(\prod_{\ell=0}^{n \mp 1} U_{\pm \mu} (\ell a \hat{\mu})\Big)}^\dagger, \quad n \equiv z/a$, where upper (lower) signs correspond to $n>0$ ($n<0$). 
\item The last step is the matching of the renormalized quasi distributions to the corresponding physical distributions, using the Large Momentum Effective Field Theory (LaMET) \cite{Xiong:2013bka}. The matching can also be performed using an intermediate step of converting the RI$'$-renormalized quasi distribution to the $\overline{MS}$ scheme and after that matching to the physical light-cone distribution.
\end{enumerate}

\begin{figure}
\vspace{-0.1cm}
\centering
\includegraphics[height=2.5cm]{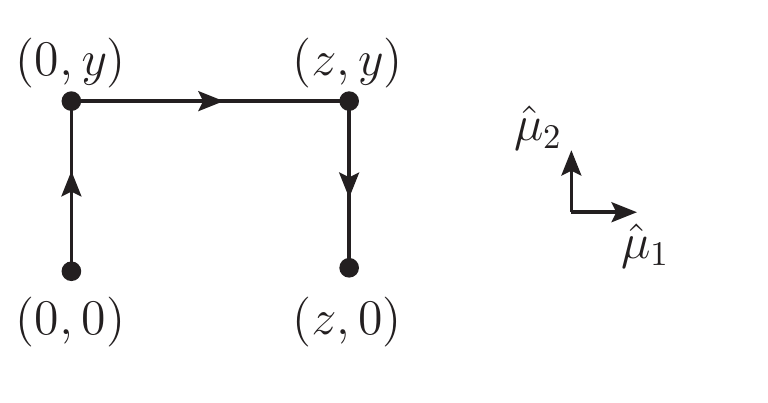}
\vspace{-0.3cm}
\caption{Staple-shaped Wilson line $W_S(z,y)$ appearing in a quasi-distribution formulation of TMDs} 
\label{fig:stable}
\end{figure}

The quasi-distribution approach can also be applied on other light-cone distributions, in which further composite Wilson-line operators are involved. For example, a staple-shaped Wilson-line operator $\overline{\psi} (0) \ \Gamma \ W_S(z,y) \psi (z)$ (see Fig. \ref{fig:stable})
is included in a quasi-distribution formulation of TMDs \cite{Musch:2010ka, Yoon:2017qzo}. In addition, the same Wilson-line operators are also involved in the pseudo-distribution approach \cite{Radyushkin:2017cyf}, which is one of the alternative approaches for extracting light-cone distribution functions on the lattice.

\section{Perturbative renormalization of Wilson-line operators}

An important issue, which needs to be addressed in order to obtain meaningful results from lattice investigations, is the renormalization of quasi-distribution functions in a fully nonperturbative manner. Given that the renormalization of a quasi-distribution function can be obtained by the renormalization of the corresponding Wilson-line operator, we study the latter in the perturbative level. Perturbative calculations can reveal possible operator mixing, which must be taken into account in the nonperturbative renormalization prescriptions. In this case, the nonlocality of Wilson-line operators combined with a chiral-symmetry breaking action lead to the appearance of mixing. Also, nonperturbative evaluations of the renormalization factors cannot be obtained directly in the $\overline{MS}$ scheme, which is typically used in phenomenology, since the definition of $\overline{MS}$ is perturbative in nature. Most naturally, one calculates them in a RI$'$-like scheme, and then introduces the corresponding conversion factors between RI$'$ and $\overline{MS}$ schemes, which rely necessarily on perturbation theory. In the present work, we investigate the renormalization of two types of Wilson-line operators: the straight-line and staple-shaped operators.

\subsection{Renormalization of straight-line operators in the presence of nonzero quark masses}

The first part of our calculation is considering the renormalization of straight-line operators in the presence of nonzero quark masses. A long write-up of our work, together with an extended list of references, can be found in Ref. \cite{Spanoudes:2018zya}. In this part, we consider a total of 16 straight-line operators of different Dirac structures:
\be 
\mathcal{O}_\Gamma^z (x) = \overline{\psi}(x) \Gamma \mathcal{P} \Big\lbrace \exp \Big(i g \int_0^z d \zeta A_\mu (x + \zeta \hat{\mu})\Big) \Big\rbrace \psi(x + z \hat{\mu}), 
\ee 
where $\Gamma = \mathbb{1},\,\gamma_5,\,\gamma_{\mu},\,\gamma_{\nu},\,\gamma_5\,\gamma_{\mu},\,\gamma_5\,\gamma_{\nu},\, \frac{1}{2} [\gamma_{\mu},\,\gamma_{\nu}],\, \frac{1}{2}[\gamma_{\nu},\,\gamma_{\rho}], \ \ (\mu \neq \nu \neq \rho \neq \mu); \ \ \mu$ and $|z|$ are the direction and the length of the straight line, respectively. The quark and antiquark fields may have different flavors: $\psi_f$ and $\overline{\psi}_{f'}$, and thus different masses. Before moving on to the presentation of our computation, let us summarize the existing knowledge on the renormalization of straight-line operators from previous studies; thus, we will give a brief progress report in this direction of research.

\subsubsection{History of renormalization of straight-line operators}

The one-loop renormalization of straight-line operators in continuum regularizations was studied many years ago. First calculations were performed in DR and the $\overline{MS}$ scheme, giving the following renormalization factor \cite{Craigie:1980qs}:
\be 
Z_\Gamma^{DR,\overline{MS}} = 1 + \frac{g^2 C_F}{16 \pi^2} \frac{3}{\varepsilon} + \mathcal{O} (g^4)
\label{ZGammaMSbar}
\ee
where $C_F = (N_c^2 - 1)/(2 N_c)$ and $N_c$ is the number of colors. According to the studies of Wilson loop operators \cite{Dotsenko:1979wb}, in regularizations other than DR, the inclusion of an additional renormalization factor of the form: $\exp (-c \ |z|/a)$, ($a:$ cut-off scale), is required in order to eliminate the linear divergences which appear, to all orders in perturbation theory.

The first perturbative lattice calculation was made recently in Ref. \cite{Constantinou:2017sej}, to one loop for massless quarks, using the Wilson/clover fermion action and a variety of Symanzik-improved gluon actions. It was shown that besides the presence of logarithmic and linear divergences, similar to those expected from the continuum, finite mixing is also present among certain pairs of the original operators under renormalization. This is deduced by comparing the lattice bare two-point amputated Green's function to the corresponding $\overline{MS}$-renormalized Green's function:
\be  
\langle \psi\,{\cal O}_{\Gamma}^z\,\bar \psi \rangle^{LR}_{amp} = \langle \psi\,{\cal O}_{\Gamma}^z\,\bar \psi \rangle^{\overline{MS}}_{amp} - \frac{g^2\,C_F}{16\,\pi^2}\, e^{i\,q_\mu z}\, \cdot \mathcal{F} \ + \ \mathcal{O} (g^4),
\ee
\be 
\mathcal{F} = \Big[\Gamma \Big(c_1 + c_2\, \beta + c_3\,\frac{|z|}{a} + \log \left(a^2 \bar\mu^2\right)\left(4-\beta\right) \Big) + \Big\{ \Gamma, \gamma_\mu \Big\} \,\Big(c_4 + c_5\,c_{\rm SW}\Big) \Big] 
\label{F}
\ee
where $\beta = 1 - \alpha$, $\alpha:$ gauge parameter, $c_{\rm SW}:$ free parameter of clover action, $\bar{\mu}:$ $\overline{\textrm{MS}}$-renormalization scale, $q$: quark external momentum, $c_i$: numerical constants and $LR$: Lattice Regularization. In Eq. \eqref{F}, along with $\Gamma_1 \equiv \Gamma$, there appear additional Dirac structures in the lattice result, $\Gamma_2 \equiv \{ \Gamma, \gamma_\mu \} /2$, which must be subtracted in the renormalization process. There are four mixing pairs in total: 
$(\Gamma_1, \Gamma_2) =$ $(\mathbb{1}, \gamma_1)$, $(\gamma_5 \gamma_2, \gamma_3 \gamma_4)$, $(\gamma_5 \gamma_3, \gamma_4 \gamma_2)$, and $(\gamma_5 \gamma_4, \gamma_2 \gamma_3)$, where by convention 1 is the direction of the straight Wilson line and 2, 3, and 4 are directions perpendicular to it.

The findings of the previous perturbative calculation on operator mixing were used in Refs. \cite{Alexandrou:2017huk, Chen:2017mzz}, for the construction of a complete nonperturbative renormalization program for the quasi-PDFs. Introducing a RI$'$ scheme which respects the observed operator-mixing pairs $(\Gamma_1, \Gamma_2)$, the renormalization factors $Z^{LR,RI'}_{(\Gamma_1, \Gamma_2)}$ for the operators $(\mathcal{O}_{\Gamma_1}, \mathcal{O}_{\Gamma_2})$ are $2\times 2$ mixing matrices:
\be
\begin{pmatrix}
\mathcal{O}_{\Gamma_1}^{LR} \\
\mathcal{O}_{\Gamma_2}^{LR}
\end{pmatrix} = (Z^{LR,RI'}_{(\Gamma_1, \Gamma_2)}) \cdot 
\begin{pmatrix}
\mathcal{O}_{\Gamma_1}^{RI'} \\
\mathcal{O}_{\Gamma_2}^{RI'} 
\end{pmatrix},
\label{Eq1}
\ee
and, in order to estimate $Z^{LR,RI'}_{(\Gamma_1, \Gamma_2)}$ nonperturbatively, one must obtain results from lattice simulations for the matrix elements of both operators. In this program, there is no need to separate the exponential renormalization factor for the elimination of the linear divergences, from the total renormalization factor, i.e.,
\be 
Z^{LR,RI'}_{(\Gamma_1, \Gamma_2)} = {\widetilde Z}^{LR,RI'}_{(\Gamma_1, \Gamma_2)}\cdot e^{- c\,|z|/a - \bar{c} |z|},
\ee
as the RI$'$ condition ensures the elimination of both linear and logarithmic divergences. Here, we note that for nonperturbative calculations, the exponential renormalization factor includes an additional term which depends on a nonperturbative scale $\bar{c}$. For the conversion of RI$'$-renormalized operators to the $\overline{MS}$ scheme, one multiplies with the appropriate $2 \times 2$ conversion factor $C^{\overline{MS},RI'}_{(\Gamma_1, \Gamma_2)}$, computed in Ref. \cite{Constantinou:2017sej} to one loop:  
\be 
Z^{LR,\overline{MS}}_{(\Gamma_1, \Gamma_2)} = {(C^{\overline{MS},RI'}_{(\Gamma_1, \Gamma_2)})}^{-1} \cdot (Z^{LR,RI'}_{(\Gamma_1, \Gamma_2)})
\label{Eq3}
\ee 
For the operators which do not exhibit any mixing, their renormalization factors are not matrices and they satisfy the standard $1 \times 1$ version of Eqs. \eqref{Eq1} - \eqref{Eq3}. 

There are also other attempts for  renormalizing the straight-line operators or directly the quasi-PDFs nonperturbatively, using alternative techniques, such as the static quark potential, the gradient flow and the auxiliary-field formalism (see references in \cite{Spanoudes:2018zya}). 

Some perturbative calculations for improving the nonperturbative renormalization programs are currently under investigation. A preliminary work by M. Constantinou and H. Panagopoulos is addressing the subtraction of lattice artifacts from nonperturbative results to one-loop level and to all orders in the lattice spacing. Also, a work of ours, which is in progress, extents the calculation of conversion factors between RI$'$ and $\overline{MS}$ schemes, to two loops. In addition, the inclusion of nonzero quark masses and their significance on the renormalization of straight-line operators have been studied by us and they are presented below. As we can conclude, the presence of quark masses affects the observed operator mixing, for both continuum and lattice regularizations, as well as the conversion factors. This study is useful for the nonperturbative investigation of heavy-quark quasi-PDFs.

\subsubsection{Calculation Setup}

\begin{figure}[thb] 
  \centering
  \includegraphics[height=1.7cm]{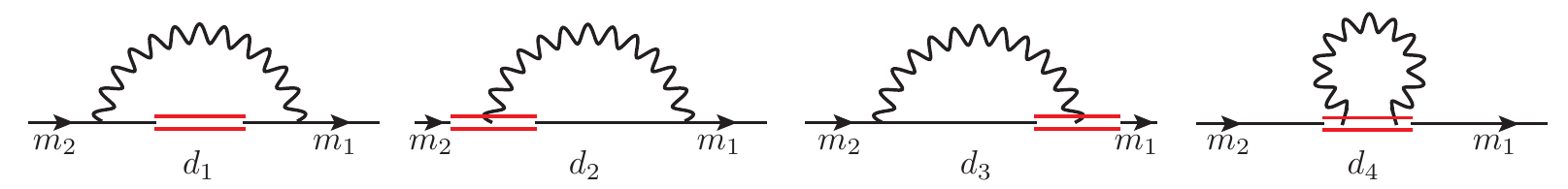}
  \caption{Feynman diagrams contributing to the one-loop calculation of the Green's functions of Wilson-line operator $\mathcal{O}_\Gamma$. The straight (wavy) lines represent fermions (gluons). The operator insertion is denoted by double straight line.}
  \label{Fig.FeynmanDiagrams}
\end{figure}

Taking into account the presence of nonzero fermion masses in our calculations, we adopt mass-dependent prescriptions for the renormalization of straight-line operators. Also, in the presence of mixing within certain subsets $(\mathcal{O}_{\Gamma_1}, \mathcal{O}_{\Gamma_2}, \ldots)$ of the original operators, their renormalization factors will have a matrix form. We define the renormalization factors which relate the bare operators $\mathcal{O}_{\Gamma_i}$, quark fields $\psi_f$ and masses $m_f^B$ with the renormalized ones via the following equations:
\be 
\mathcal{O}^Y_{\Gamma_i} = {\Big[{(Z^{X,Y}_{(\Gamma_1, \Gamma_2, \ldots)})}^{-1}\Big]}_{ij} \mathcal{O}_{\Gamma_j}, \qquad \psi_f^Y = {(Z^{X,Y}_{\psi_f})}^{-1/2} \psi_f, \qquad m_f^Y ={(Z^{X,Y}_{m_f})}^{-1} m_f^B,
\ee
where $X (Y)$ stands for the regularization (renormalization) scheme: $X = DR, LR, \ldots$, $Y = \overline{MS}, RI', \ldots$. We note that for regularizations which break chiral symmetry (such as Wilson/clover fermions), an additive mass renormalization is also needed, beyond one loop ($m^B = m_0 - m_c$, where $m_0 (m_c)$ is the Lagrangian (additive/critical) mass). However, this is irrelevant for our one-loop calculations. For the calculation of $Z^{X,Y}_{(\Gamma_1, \Gamma_2, \ldots)}$, we need the evaluation of the corresponding bare two-point amputated  Green's functions ${\langle\psi_f\,{\mathcal O}_{\Gamma_i}\,\bar \psi_{f'} \rangle}^X_{\rm amp}$, which are related to the renormalized ones via:
\be 
{\langle\psi_f^Y\,{\mathcal O}^Y_{\Gamma_i}\,\bar \psi_{f'}^Y \rangle}_{\rm amp} = {(Z^{X,Y}_{\psi_f})}^{1/2} \, {(Z^{X,Y}_{\psi_{f'}})}^{1/2} \, {\Big[{(Z^{X,Y}_{(\Gamma_1, \Gamma_2, \ldots)})}^{-1}\Big]}_{ij} {\langle\psi_f\,{\mathcal O}_{\Gamma_j}\,\bar \psi_{f'} \rangle}^X_{\rm amp}
\label{Greensfunction}
\ee
In the massive case, renormalization factors of the fermion and antifermion fields appearing in bilinear operators of different flavor content may differ among themselves, as the fields have generally different masses. There are four one-loop Feynman diagrams corresponding to ${\langle\psi_f\,{\mathcal O}_{\Gamma_i}\,\bar \psi_{f'} \rangle}^X_{\rm amp}$, shown in Fig. \ref{Fig.FeynmanDiagrams}. 
The last diagram ($d_4$) does not depend on the quark masses, and therefore its contribution is the same as that of the massless case.

The calculation of Green's functions of straight-line operators demonstrates that additional contributions of different Dirac structures from that of the original operator are presented for each operator:
\begin{gather}
{\langle\psi_f\,{\mathcal O}_{\Gamma_1}\,\bar \psi_{f'} \rangle}_{\rm amp}^{\text{X, 1-loop}} = g^2 \Big(\lambda_{11} {\langle\psi_f\,{\mathcal O}_{\Gamma_1}\,\bar \psi_{f'} \rangle}_{\rm amp}^{\text{tree}} + \lambda_{12} {\langle\psi_f\,{\mathcal O}_{\Gamma_2}\,\bar \psi_{f'} \rangle}_{\rm amp}^{\text{tree}} + \ldots\Big) \nonumber \\
{\langle\psi_f\,{\mathcal O}_{\Gamma_2}\,\bar \psi_{f'} \rangle}_{\rm amp}^{\text{X, 1-loop}} = g^2 \Big(\lambda_{21} {\langle\psi_f\,{\mathcal O}_{\Gamma_1}\,\bar \psi_{f'} \rangle}_{\rm amp}^{\text{tree}} + \lambda_{22} {\langle\psi_f\,{\mathcal O}_{\Gamma_2}\,\bar \psi_{f'} \rangle}_{\rm amp}^{\text{tree}} + \ldots\Big) \nonumber \\
\vdots \nonumber 
\end{gather}
where ${\langle\psi_f\,{\mathcal O}_{\Gamma_i}\,\bar \psi_{f'} \rangle}_{\rm amp}^{\text{tree}} = \Gamma_i \exp (i q_\mu z)$. These contributions are not divergent and thus, they are irrelevant for the renormalization factors in DR and the $\overline{MS}$ scheme. However, some of these contributions are regularization-dependent, which means that they must be removed in a RI$'$-like scheme. Of course, this cannot be achieved by a simple multiplicative renormalization, but we need to introduce mixing matrices for certain subsets of operators. The results of the present work demonstrate that the presence of quark masses affects the observed operator-mixing pairs, due to the chiral-symmetry breaking of mass terms in the fermion action. Compared to the massless case on the lattice, the mixing pairs remain the same for operators with equal masses of external quark fields, i.e., $(\mathbb{1}, \gamma_1)$, $(\gamma_5 \gamma_2, \gamma_3 \gamma_4)$, $(\gamma_5 \gamma_3, \gamma_4 \gamma_2)$, and $(\gamma_5 \gamma_4, \gamma_2 \gamma_3)$, where by convention 1 is the direction of the straight Wilson line and 2, 3, and 4 are directions perpendicular to it. However, for operators with different masses of external quark fields, flavor-symmetry breaking leads to four additional mixing pairs: $(\gamma_5, \gamma_5 \gamma_1)$, $(\gamma_2,\gamma_1 \gamma_2)$, $(\gamma_3,\gamma_1 \gamma_3)$, and $(\gamma_4,\gamma_1 \gamma_4)$. Of course, the operator mixing depends on the precise regularization-independent definition of the RI$'$ scheme.

There is, a priori, wide flexibility in defining RI$'$-like normalization conditions for Green's functions. Given that there is a residual rotational (or hypercubic, on the lattice) symmetry with respect to the three transverse directions to the straight-line, including also reflections, no mixing needs to occur among operators which do not transform in the same way under this residual symmetry. Therefore, it is natural to adopt the minimal prescription of the following renormalization condition which respects the observed operator mixing pairs, mentioned above:
\begin{align}
{\rm tr}\Big[\Big({\langle\psi_f^{\text{RI}'}\,{\mathcal O}_{\Gamma_i}^{\text{RI}'}\,\bar \psi_{f'}^{\text{RI}'} \rangle}_{\rm amp}\Big) \Big({\langle\psi_f\,{\mathcal O}_{\Gamma_j}\,\bar \psi_{f'} \rangle}_{\rm amp}^{\text{tree}}\Big)^\dagger\Big] \Bigg|_{
\begin{smallmatrix}
q_\nu = \bar{q}_\nu
\end{smallmatrix}
} \! \! \! \! \! &= \, \nonumber \\ 
{\rm tr}\Big[\Big({\langle\psi_f\,{\mathcal O}_{\Gamma_i}\,\bar \psi_{f'} \rangle}_{\rm amp}^{\text{tree}}\Big) \Big({\langle\psi_f\,{\mathcal O}_{\Gamma_j}\,\bar \psi_{f'} \rangle}_{\rm amp}^{\text{tree}}\Big)^\dagger\Big] \Bigg|_{
\begin{smallmatrix}
q_\nu = \bar{q}_\nu
\end{smallmatrix}
} \! \! \! \! \! &= \, \ \ 4 N_c \delta_{ij}, 
\label{RIcondition}
\end{align}
where $i,j = 1,2$ and $\bar{q}_\nu$ is the $RI'$-renormalized four-vector scale. 

We also include the mass-dependent RI$'$ renormalization conditions for the quark fields and masses, in textbook fashion:
\begin{align}
&\textrm{tr} \Big[- \frac{i \slashed{q}}{q^2} {\left(\langle\psi^{\text{RI}'}_f \bar{\psi}^{\text{RI}'}_f\rangle\right)}^{-1}\Big] \Bigg|_{
\begin{smallmatrix}
q_\nu = \bar{q}_\nu
\end{smallmatrix}
} \! \! \! \! \! = \, \textrm{tr} \Big[- \frac{i \slashed{q}}{q^2} {\Big(\langle\psi_f \bar{\psi}_f\rangle^{\text{tree}}\Big)}^{-1}\Big] \Bigg|_{
\begin{smallmatrix}
q_\nu = \bar{q}_\nu
\end{smallmatrix}
} \! \! \! \! \! = \, 4 N_c 
\end{align}
\begin{align}   
&\textrm{tr} \Big[\mathbb{1} {\left(\langle\psi^{\text{RI}'}_f \bar{\psi}^{\text{RI}'}_f\rangle\right)}^{-1}\Big] \Bigg|_{
\begin{smallmatrix}
q_\nu = \bar{q}_\nu
\end{smallmatrix}
} \! \! \! \! \! = \, 4 N_c \ m_f^{\text{RI}'} 
\end{align} 

As a consequence of the operator-pair mixing, the conversion factors between RI$'$ and $\overline{MS}$ schemes will be $2 \times 2$ mixing matrices. Being regularization independent, they can be evaluated more easily in DR. They are defined as: 
\be
\Big[C_{(\Gamma_1, \Gamma_2)}^{\overline{MS},{\rm RI}'}\Big]_{ij} = \sum_{k=1}^2 \Big[(Z_{(\Gamma_1, \Gamma_2)}^{DR,\overline{MS}})^{-1}\Big]_{ik} \cdot \Big[Z_{(\Gamma_1, \Gamma_2)}^{DR,{\rm RI}'}\Big]_{kj} = \sum_{k=1}^2 \Big[(Z_{(\Gamma_1, \Gamma_2)}^{LR,\overline{MS}})^{-1}\Big]_{ik} \cdot \Big[Z_{(\Gamma_1, \Gamma_2)}^{LR,{\rm RI}'}\Big]_{kj},
\label{ConvFactors}
\ee
where \ $Z_{(\Gamma_1, \Gamma_2)}^{DR,\overline{MS}}$ is diagonal because there is no mixing in $(DR, \overline{MS})$.

\subsubsection{Calculation and Results}

For the calculation of the momentum-loop integrals in the Green's functions in DR, we follow the standard procedure of introducing Feynman parameters. The presence of an exponential function in the loop-integrals leads to a generalization of the standard formulas for the computation of Feynman parameter integrals. The resulting formulas depend on modified Bessel functions of second kind $K_\nu (z)$: 
\be 
I (\alpha) \equiv \int \frac{d^D p}{(2 \pi)^D} \frac{e^{i p_\mu z}}{{(p^2 + 2 \ k \cdot p + m^2)}^{\alpha}} = \frac{2^{1-\alpha-D/2} \ {\vert z \vert}^{\alpha - D/2} \ e^{-i k_{\mu} z}}{\pi^{D/2} \ \Gamma (\alpha) \ (m^2 - k^2)^{\alpha/2 - D/4}} \ K_{-\alpha + D/2} (\sqrt{m^2-k^2} \ \vert z \vert), 
\ee
\be 
\int \frac{d^D p}{(2 \pi)^D} \frac{e^{i p_\mu z} \ p_{\nu_1} \cdots p_{\nu_n}}{{(p^2 + 2 \ k \cdot p + m^2)}^{\alpha}} = \frac{{(-1)}^n \ \Gamma (\alpha - n)}{2^n \ \Gamma (\alpha)} \frac{\partial}{\partial k_{\nu_1}} \cdots \frac{\partial}{\partial k_{\nu_n}} I (\alpha - n) 
\ee

Our results for the bare Green's functions in DR (see full expressions in Ref. \cite{Spanoudes:2018zya}) demonstrate that UV-divergent terms stem only from diagrams $d_2 - d_4$. These terms are multiples of the tree-level values of Green's functions and therefore do not lead to any mixing. The finite terms of Green's functions are complex functions and they exhibit a nontrivial dependence on the dimensionless quantities: $z q_\mu$, $z m_i$, $(i=1,2)$, besides the standard logarithmic dependence: $\log{({\bar{\mu}}^2 / q^2)}$. Also, our results are not analytic functions of z near 0; this was expected due to the appearance of contact terms beyond tree level. 

Our results on the renormalization factors of straight-line operators in (DR, $\overline{MS}$) are in agreement with previous studies (see Eq. \eqref{ZGammaMSbar}). As we expected, they are independent of fermion masses, Wilson-line length and gauge parameter. This independence is also valid to all loops, since the most dominant pole at every loop cannot depend on the fermion masses or on the external momenta or on the renormalization scale, thus there is no dimensionless z-dependent factor that could appear in the pole part. Also, $\overline{MS}$ guarantees gauge invariance, since it removes only universal divergences.

In contrast to the massless case, the renormalization factors in (DR, RI$'$) and thus the conversion factors between RI$'$ and $\overline{MS}$ are nondiagonal matrices. However, in the case of equal quark masses, the nondiagonal matrix elements of both renormalization and conversion factors vanish for the pairs $(\gamma_5, \gamma_5 \gamma_\mu)$, $(\gamma_\nu, \gamma_\mu \gamma_\nu)$, $(\nu \neq \mu)$. Just to give an example, we present a matrix element of one of the conversion factors:
\be
\begin{split}
\left[C^{\overline{MS},RI'}_{S,V_{\mu}} \right]_{12} = \ \frac{g^2 C_F}{16 \pi ^2} \Bigg\{&-\beta z \Big( m_1 f_{21} + m_2 f_{22} \Big) - \beta \bar{q}_{\mu}^2 \Big[m_1 \Big(g_{31} - z f_{31}\Big)+ m_2 \Big(g_{32} - z f_{32}\Big) \Big] \\
&+ i \beta \bar{q}_{\mu} \Big[ m_1 (m_1^2 + \bar{q}^2) g_{41} + m_2 (m_2^2 + \bar{q}^2) g_{42} - (\bar{q}^2 - \bar{q}_{\mu}^2) \Big(m_1 g_{51} + m_2 g_{52}\Big)\Big] \\
&+ 2 \Big(m_1 g_{11}+m_2 g_{12}\Big) + \Big(m_1+m_2\Big) \Bigg[ (\beta +2) z h_1  - i \beta  \bar{q}^2 \left| z\right| \bar{q}_{\mu} h_8 \\
&+ i \left| z\right| \bar{q}_{\mu} \Big((\beta - 2) h_5 + 2 h_6\Big) +\frac{1}{2} i \beta  z^2 \bar{q}_{\mu} \Big(h_2-\bar{q}^2 h_3\Big) -\beta  \bar{q}^2 z \left| z\right| \Big(h_5-h_6-h_7\Big) \Bigg] \Bigg\}\\ 
&\hspace{-1cm} + \mathcal{O} (g^4),  
\end{split} 
\ee
where $S (V_\mu)$ denotes the Dirac matrix $\mathbb{1} (\gamma_\mu)$, and $f_{ij}$, $g_{ij}$, $h_i$ denote integrals of modified Bessel functions, over Feynman parameters. To give a flavor of what these integrals look like, we provide an example for each type:
\be  
f_{31} = \int_0^1 dx \ \exp{\left(- i \bar{q}_{\mu} x z \right)} \ K_0\left( \left| z\right| s \right) \ (1 - x) \ \frac{x^2}{{s}^2},
\ee
\be
g_{31} = \int_0^1 dx \ \int_0^z d\zeta \ \exp{\left(- i \bar{q}_{\mu} x \zeta \right)} \ K_0\left( \left| \zeta \right| s \right) \ (1 - x) \ \frac{x^2}{s^2},
\ee
\be
h_3 = \int_0^1 dx_1 \ \int_0^{1 - x_1} \hspace{-0.25cm}dx_2 \ \exp{\left(- i \bar{q}_{\mu} (x_1 + x_2) z \right)} \ K_0\left( \left| z\right| t \right) \ (1 - x_1 - x_2) \ \frac{(x_1 + x_2)^2}{t^2},
\ee
where $s \equiv {\Big[\bar{q}^2 \left(1-x\right) x + m_1^2 x\Big]}^{1/2},$ and $t \equiv {\Big[ \bar{q}^2 \left(1-x_1 - x_2\right) \left(x_1 + x_2\right) + m_1^2 \ x_1 + m_2^2 \ x_2 \Big]}^{1/2}$. These integrals do not have a closed analytic form, but they are convergent and can be computed numerically. The expressions for the remaining conversion factors' matrix elements are written out explicitly in Ref. \cite{Spanoudes:2018zya}, along with the definitions of the corresponding integrals over modified Bessel functions.

\begin{figure}
\centering
\includegraphics[height=3.8cm]{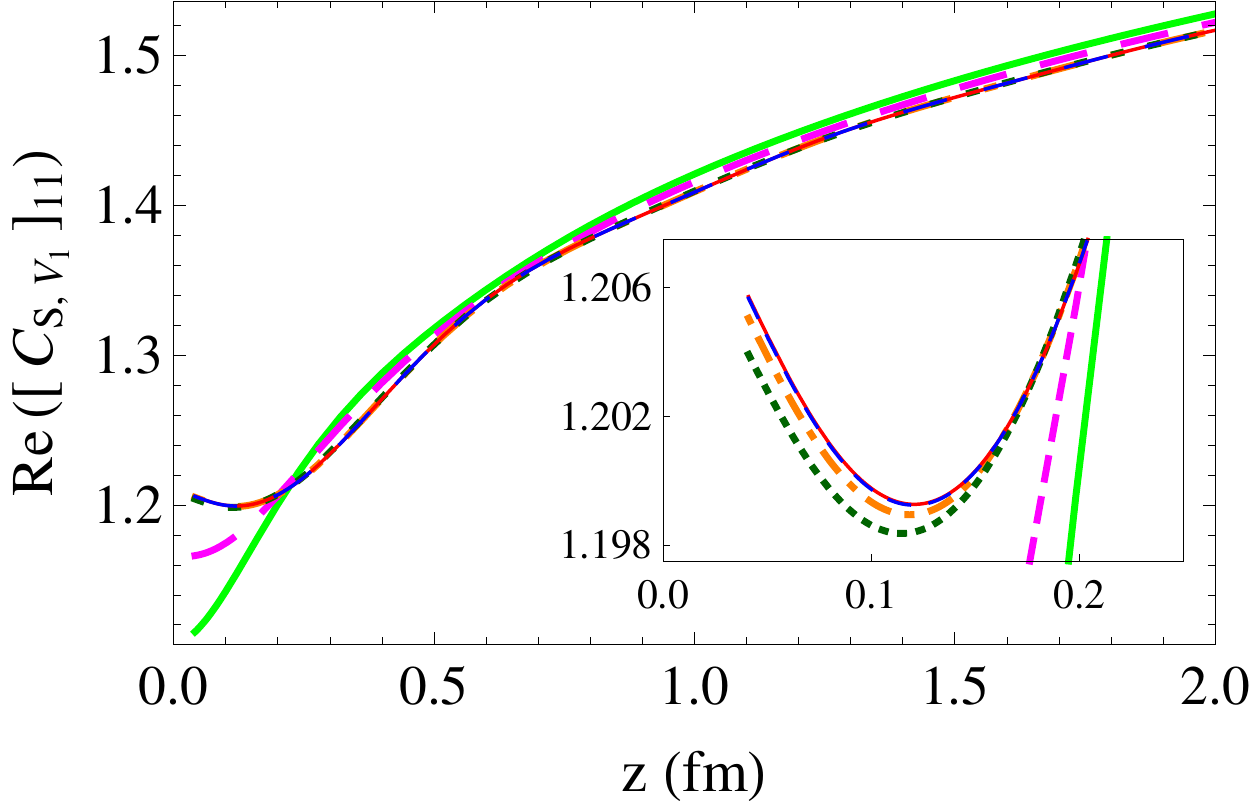} \hspace{0.2cm}
\includegraphics[height=3.85cm]{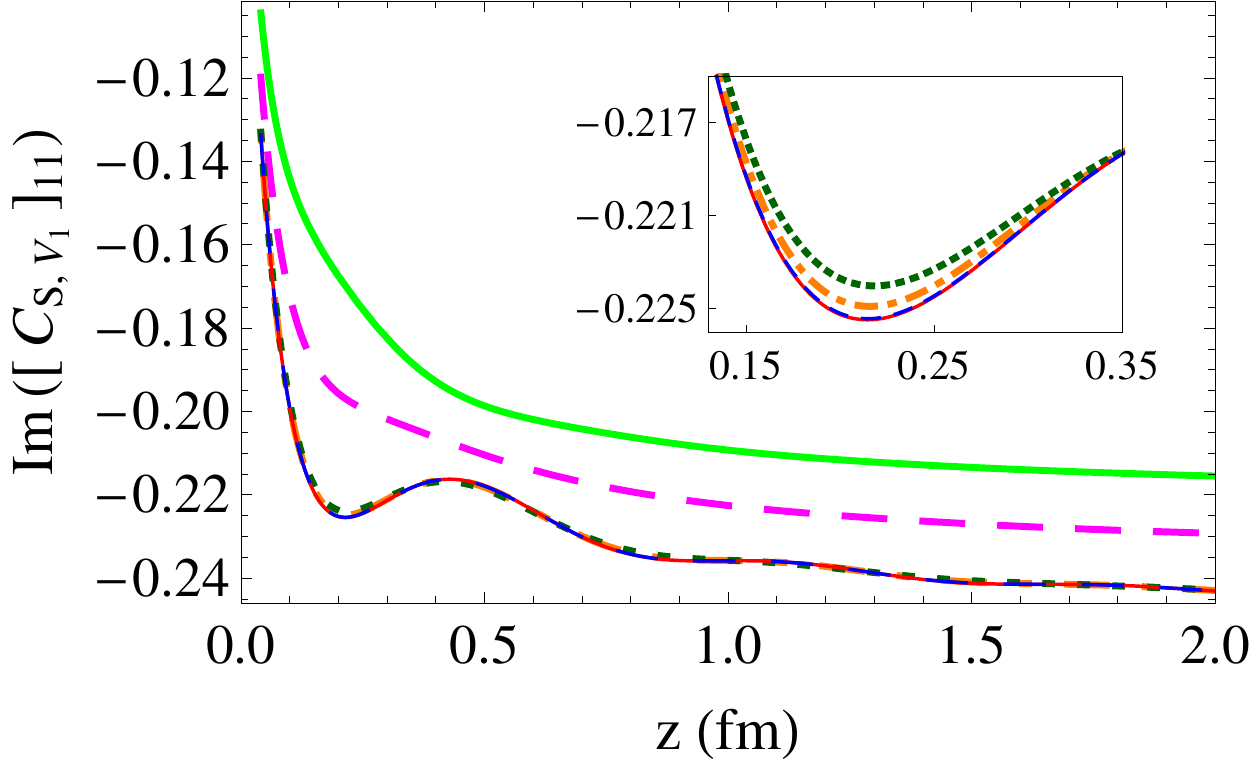} \\ 
\includegraphics[height=3.8cm]{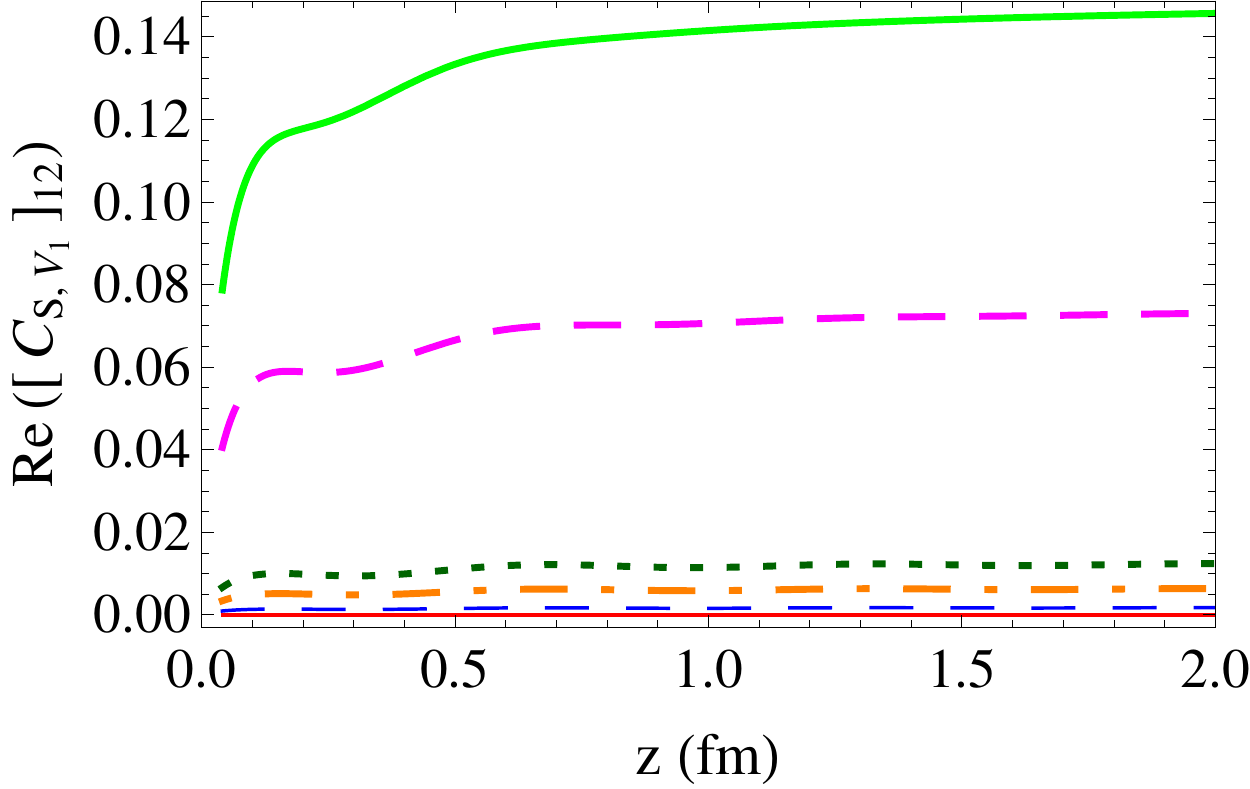} \hspace{0.2cm}
\includegraphics[height=3.85cm]{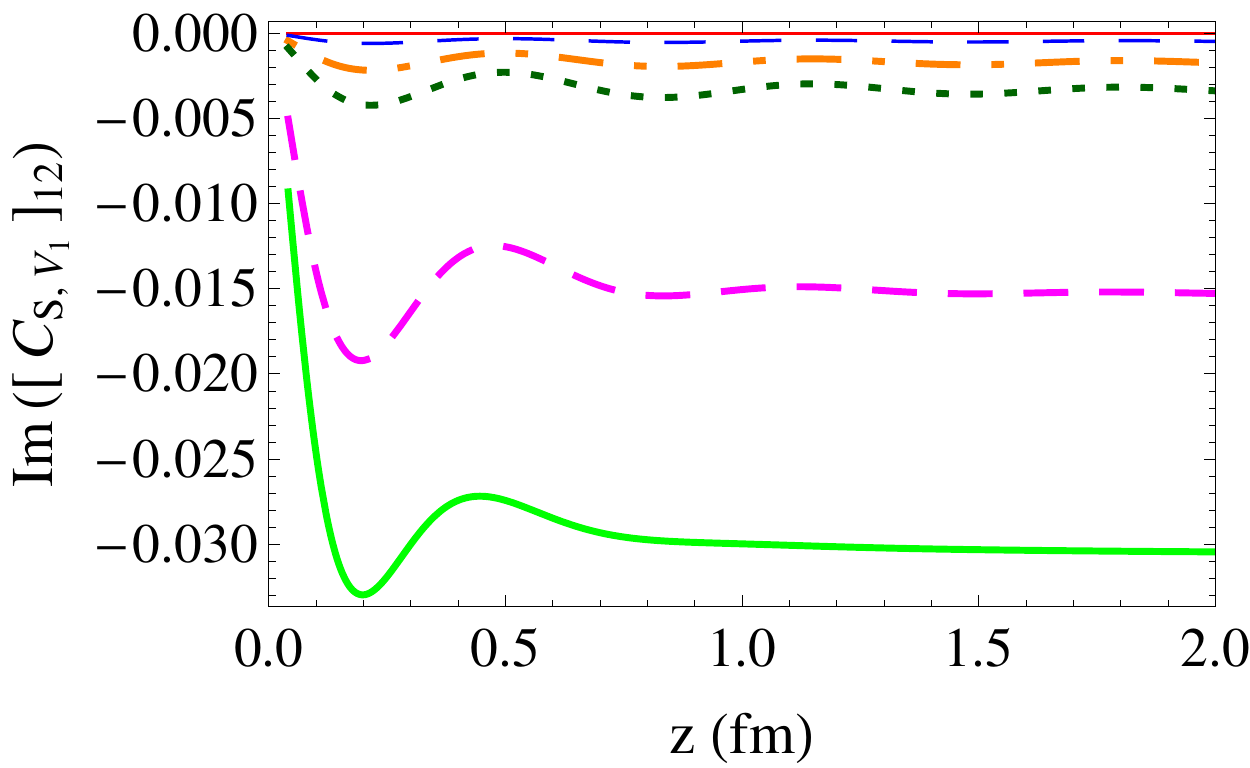} \\ 
\includegraphics[height=1.35cm]{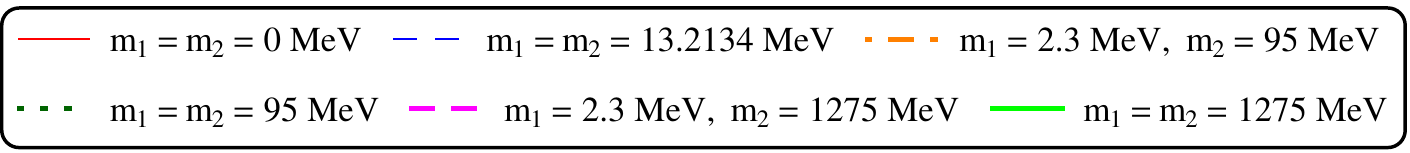} 
\caption{Real (left panels) and imaginary (right panels) parts of the conversion factor matrix elements (1,1) and (1,2) for the operator pair ($S$, $V_1$) as a function of z, for different values of quark masses.}
\label{plot}
\end{figure}

In Fig. \ref{plot}, we plot the real and imaginary parts of a representative diagonal and nondiagonal conversion factor matrix element as a function of Wilson-line length z, for different cases of external quark masses. We select certain values of the free parameters used in ETMC simulations: $g^2 = 3.077$, $\beta = 1$, $C_F = \frac{4}{3}$, $\bar{\mu} = 2$ GeV, $\bar{q} = \frac{2 \pi}{32 a} (\frac{n_t}{2} + \frac{1}{4},0,0,n_z)$, $a = 0.082$ fm, $n_t = 8$, $n_z = 4$. We note that the plots can be easily extended for negative values of z, since the real part of a diagonal (nondiagonal) element is an even (odd) function of z, while the imaginary part is odd (even). 
We observe that the real parts are an order of magnitude larger than the imaginary parts. Also, the nondiagonal element is an order of magnitude smaller than the diagonal element. Furthermore, for large values of z, the nondiagonal element, as well as the imaginary part of the diagonal element, tend to stabilize, while the real part of the diagonal element tends to increase. Thus, a two-loop calculation would be certainly welcome. Comparing the six cases of quark masses, we deduce that the impact of mass becomes significant when we include a strange or a charm quark. Therefore, we conclude that for external quarks lighter than strange, we may ignore the mass terms in the calculations of quasi-PDFs, while for heavier quarks, the mass terms lead to perceptible changes. More graphs can be found in Ref. \cite{Spanoudes:2018zya}.

\subsection{Renormalization of staple-shaped operators}

The second part of our calculation addresses the renormalization of staple-shaped operators in both continuum (DR) and lattice (Wilson/clover fermions and Symanzik improved gluons) regularizations. The difference between these operators and the straight-line operators is the shape of the involved Wilson line, which affects the UV divergences, as well as the operator-mixing pairs. We will highlight some results of our preliminary work (together with M. Constantinou) on the renormalization of these operators. Our findings can be a guidance to the nonperturbative renormalization of quasi-TMDs. 

\subsubsection{Preliminary results}

\begin{figure}
\vspace{-0.1cm}
\centering
\includegraphics[height=1.7cm]{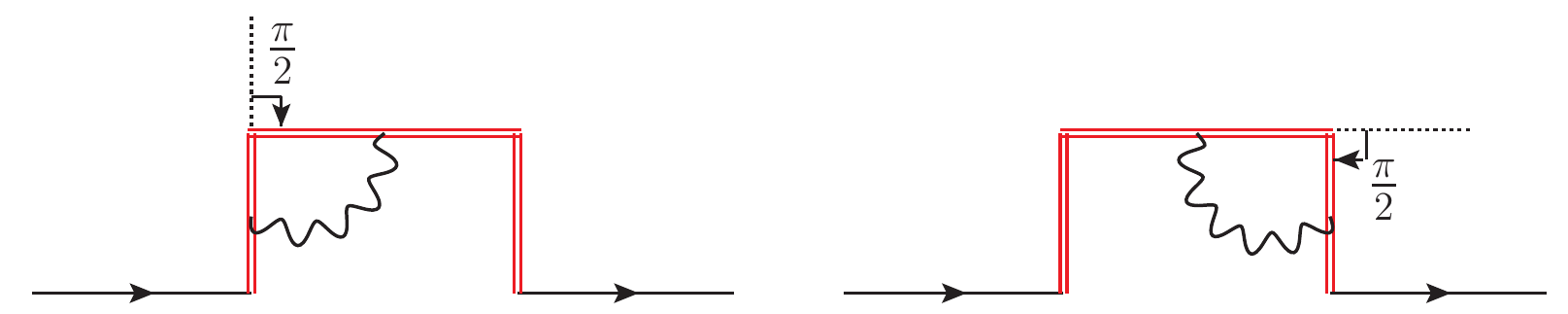} 
\vspace{-0.1cm}
\caption{A staple-shaped Wilson line has cusps of angle $\pi / 2$.}
\label{cusp_pi2}
\end{figure}

We define the staple-shaped operators as:
\be
\mathcal{O}_\Gamma^{z, \ y} (x) = \bar\psi(x) \ \Gamma \ W(x,x+y \hat{\mu}_2,x+y \hat{\mu}_2+z \hat{\mu}_1,x+z \hat{\mu}_1) \ \psi(x + z {\hat{\mu}}_1),  
\ee
where 
\begin{align}
W(x,x+y \hat{\mu}_2,&x+y \hat{\mu}_2+z \hat{\mu}_1,x+z \hat{\mu}_1) = \nonumber \\
& \mathcal{P} \Big\{ \Big(e^{i g \int_0^y d \bar{\zeta} A_{\mu_2} (x + \bar{\zeta} {\hat{\mu}}_2)}\Big) \cdot \Big(e^{i g \int_0^z d \zeta A_{\mu_1} (x + y + \zeta {\hat{\mu}}_1)}\Big) \cdot \Big( e^{i g \int_0^y d \bar{\zeta} A_{\mu_2} (x + z + \bar{\zeta} {\hat{\mu}}_2)} \Big)^\dagger \Big\}. 
\end{align} 
The presence of cusps in the staple-shaped Wilson line (see Fig. \ref{cusp_pi2}) 
leads to additional divergences in Green's functions, compared to those of straight-line operators, which depend on the cusp angles ($\pi / 2$). Our result for the renormalization factors of operators in (DR, $\overline{MS}$) is given by:
\be
Z_\Gamma^{DR,\overline{MS}} = 1 + \frac{g^2 C_F}{16 \pi^2} \frac{7}{\varepsilon} + \mathcal{O} (g^4).
\ee
The term $\frac{g^2 C_F}{16 \pi^2} \frac{7}{\varepsilon}$ comes from the sum of straight-line divergences (three straight-line segments): $3 (\frac{g^2 C_F}{16 \pi^2} \frac{3}{\varepsilon})$ and cusp divergences (two cusps of angle $\pi / 2$): $2(-\frac{g^2 C_F}{16 \pi^2} \frac{1}{\varepsilon})$, which agree with previous studies of cusp divergences (e.g., \cite{Brandt:1981kf}).

On the lattice, our results demonstrate that, as we expected, there is a linear divergence which depends on the length of the staple line ($|z| + 2 |y|$). Also, comparing to the $\overline{MS}$-renormalized Green's functions we found that operator-mixing pairs occur which are totally different from those of the straight-line operators:
\be
\langle \psi\,{\cal O}_{\Gamma}^{z, \ y} \,\bar \psi \rangle^{LR} = \langle \psi\,{\cal O}_{\Gamma}^{z, \ y}\,\bar \psi \rangle^{\overline{MS}} - \frac{g^2\,C_F}{16\,\pi^2}\, e^{i\,q_{\mu_1} z}\, \cdot \mathcal{F} \ + \ \mathcal{O} (g^4),
\ee
\be
\mathcal{F} = \Big[\Gamma \Big(c_1 + c_2 \, \beta + c_3\,\frac{|z| + 2 \ |y|}{a} + \log \left(a^2 \bar\mu^2\right) \left(8-\beta\right) \Big) + \text{sgn}(y) \Big[ \Gamma, \gamma_{\mu_2} \Big] \,\Big(c_4 + c_5 \,c_{\rm SW}\Big) \Big]
\ee
where $c_2 = 3.79201(1)$ and the remaining $c_i$, are also numerical constants which depend on the gluon action in use. In particular, the mixing pairs are: $(\gamma_5, \gamma_5 \gamma_2)$, $(\gamma_1, \gamma_1 \gamma_2)$, $(\gamma_3, \gamma_3 \gamma_2)$, $(\gamma_4, \gamma_4 \gamma_2)$, where by convention: ($\mu_1 = 1$, $\mu_2 = 2$) specifies the plane on which the staple lies, and 3, 4 are directions perpendicular to this plane. More details of our calculation and more results, including the numerical values of $c_i$, will be presented in a forthcoming publication \cite{Spanoudes:2018}. 

\section{Follow-up work}

\vspace{-0.08cm}There are many things to be done in the perturbative studies of Wilson-line operators. The calculation of two-loop conversion factors, as well as the evaluation of lattice artifacts and their subtraction from the nonperturbative estimates, are already in progress. A possible extension is the addition of stout smearing of gluon links in the fermion action, and also in the definition of Wilson-line operators. Finally, our perturbative analysis can also be applied to the study of further composite Wilson-line operators, relevant to different quasi-distribution functions, e.g., quasi-GPDs, gluon quasi-PDFs, etc. \\
\vspace{-0.1cm} \\
\textbf{Acknowledgements:} We would like to thank our collaborator M. Constantinou for useful discussions. G.S. acknowledges financial support by the University of Cyprus, within the framework of Ph.D. student scholarships.

\vspace{-0.1cm}
\bibliographystyle{JHEP}                     
\bibliography{references}

%

\end{document}